\documentstyle[twocolumn,aps,prl,epsf]{revtex}

\begin{document}

\title{ENERGY LANDSCAPES, SUPERGRAPHS, AND ``FOLDING FUNNELS'' IN SPIN SYSTEMS}

\author{Piotr Garstecki,$^1$ Trinh Xuan Hoang,$^2$ and Marek Cieplak$^2$}

\address{$^1$Institute of Chemical Physics, Polish Academy of Sciences,
ul. Kasprzaka 44/52, 01-224 Warsaw, Poland}
\address{$^2$Institute of Physics, Polish Academy of Sciences,
Al. Lotnikow 32/46, 02-668 Warsaw, Poland }
\address{
\centering{
\medskip\em
{}~\\
\begin{minipage}{14cm}
Dynamical connectivity graphs, which describe dynamical transition
rates between local energy minima of a system, can be displayed
against the background of a disconnectivity graph which represents the
energy landscape of the system. The resulting supergraph describes
both dynamics and statics of the system in a unified coarse-grained
sense. We give examples of the supergraphs for several two dimensional
spin and protein-related systems. We demonstrate that disordered
ferromagnets have supergraphs akin to those of model proteins whereas
spin glasses behave like random sequences of aminoacids which fold
badly.
{}~\\
{}~
\end{minipage}
}}

\maketitle
\newpage

\section{INTRODUCTION}

The concept of energy landscapes has played a significant role in
elucidating  kinetics of protein folding \cite{Bryngelson,Wolynes}.
An energy landscape can be visualized by using the so called
disconnectivity graphs \cite{Becker} that show patterns of pathways
between the local energy minima of a system.  A pathway consists of
consecutive moves that are allowed kinetically. The pathways indicated
in a disconnectivity graph are selected to be those which provide a
linkage at the lowest energy cost among all possible trajectories
between two destinations.  Thus, at each predetermined value of a
threshold energy, the local energy minima are represented as divided
into disconnected sets of minima which are mutually accessible through
energy barriers.  The local minima which share the lowest energy
barrier are joined at a common node and are said to be a part of a
basin corresponding to the threshold.

The disconnectivity graphs have proved to be useful tools to elucidate
the energy landscape of a model of a short peptide \cite{Becker} and
of several simple molecular systems. In particular, Wales, Miller, and
Walsh \cite{Wales} have constructed disconnectivity graphs for the
archetypal energy landscapes of a cluster of 38 Lennard-Jones atoms,
the molecule of C$_{60}$, and 20 molecules of water.  The work on the
Lennard-Jones systems has been recently extended by Doye {\it et al.}
\cite{Doye}. 
The graph for a well folding protein is expected to have an appearance
of a ``palm tree.'' This pattern has a well developed basin of the
ground state and it also displays several branches to substantially
higher lying local energy minima.  Such a structure seems naturally
associated with the existence of a folding funnel.  The atomic level
studies of the 4-monomer peptide considered by Becker and Karplus
\cite{Becker} yield a disconnectivity graph which suggests that this
expected behavior may be correct. Bad folders are expected to have
disconnectivity graphs similar to either a ``weeping willow'' or a
``banyan tree'' \cite{Becker,Wales} in which there are many competing
low lying energy minima.

We accomplish several tasks in this paper. The first of these, as
addressed in Sec. II, is to construct disconnectivity graphs for two
lattice heteropolymers the dynamics of which have been already studied
exactly \cite{Malte}. One of them is a model of a protein, in the
sense that it has excellent folding properties, 
and we shall refer to it as a good folder. The other has very poor
folding properties, i.e. it is a bad
folder and is thus a model of a random sequence of aminoacids. We show
that, indeed, only the good folder has a protein-like disconnectivity
graph. 

In Sec. III we study the archetypal energy landscapes corresponding
to small two dimensional (2D) Ising spin systems with the
ferromagnetic and spin glassy exchange couplings.  We demonstrate that
disordered ferromagnets have protein-like disconnectivity graphs
whereas spin glasses behave like bad folders. This is consistent with
the concept of minimal structural frustration \cite{frustr}, or
maximal compatibility, that has been introduced to explain why natural
proteins have properties which differ from those characterizing random
sequences of aminoacids. It is thus expected that spin systems which
have the minimal frustration in the exchange energy, i.e.  the
disordered ferromagnets, would be the analogs of proteins.  In fact,
we demonstrate that the kinetics of ``folding,'' i.e. the kinetics of
getting to the fully aligned ground state of the ferromagnet by
evolving from a random state, depends on temperature, $T$, the way a
protein does. Finding a ground state of a similarly sized spin glass
takes place significantly longer.

The disconnectivity graphs characterize the phase space of a system
and, therefore, they relate primarily to the equilibrium properties --
the dynamics is involved only through a definition of what kinds of
moves are allowed, but their probabilities of being implemented are of
no consequence.  Note that even if the disconnectivity graph indicates
a funnel-like structure, the system may not get there if the
temperature is not right. Thus a demonstration of the existence of a
funnel must involve an actual dynamics. In fact, another kind of
connectivity graphs between local energy minima has been introduced
recently precisely to describe the $T$-dependent dynamical linkages
\cite{coarse} in the context of proteins.  We shall use the phrase a
``dynamical connectivity graph'' to distinguish this concept from that
of a ``disconnectivity graph'' of Becker and Karplus.  The idea behind
the dynamical connectivity graphs is rooted in a coarse grained
description of the dynamics through mapping of the system's
trajectories to underlying effective states. In ref. \cite{coarse},
the effective states are the local energy minima arising as a result
of the steepest descent mapping. In ref. \cite{cells}, the steepest
descent procedure is followed by an additional mapping to a closest
maximally compact conformation. The steepest descent mapping has been
already used to describe glasses \cite{Stillinger} and spin glasses
\cite{Jaeckle} in terms of their inherent, or hidden, valley
structures.

In the dynamical connectivity graphs, the linkages are not uniform in
strength. Their strengths are defined by the frequency with which the
two effective states are visited sequentially during the temporal
evolution. The strengths are thus equal to the transition rates and
they vary significantly from linkage to linkage and as a function of
$T$. An additional characteristic used in such graphs is the fraction
of time spent in a given effective state, without making a transition.
This can be represented by varying sizes of symbols associated with
the state.

In the context of these developments, it seems natural to combine the
two kinds of coarse-graining graphs, equilibrium and dynamical, into
single entities -- the supergraphs. Such supergraphs can be
constructed by placing  the information about the $T$-dependent
dynamical linkages on the energy landscape represented by the
disconnectivity graph.  This procedure is illustrated in Sec. IV for
the case of the two heteropolymers discussed in Sec. II.  The
procedure is then applied to selected spin systems. In each case,
knots of significant dynamical connectivities within the ground state
basin develop around a temperature at which the specific heat has a
maximum.  These knots disintegrate on lowering the $T$ if the system
is a spin glass or a bad folder. For good folders and non-uniform
ferromagnets the dynamical linkages within the ground state basin
remain robust.

We hope that this kind of combined characterization, by the
supergraphs, of both the dynamics and equilibrium pathways existing in
many body systems might prove revealing also in the case of other
systems, e.g., such as the molecular systems considered in ref.
\cite{Wales}.

\section{ENERGY LANDSCAPES IN 2D LATTICE PROTEINS}

Lattice models of heteropolymers allow for an exact determination of the
native state, i.e. of the ground state of the system, and are endowed with
a simplified dynamics. These two features have allowed for significant
advancement in understanding of protein folding \cite{Dill}.

Here, we consider two 12-monomer sequences of model heteropolymers, $A$ and
$B$, on a two-dimensional square lattice. These sequences have been
defined in terms of Gaussian contact energies (the mean equal to $-1$
and the dispersion to 1, roughly) in ref. \cite{Malte}. They have been
studied \cite{Malte,coarse} in great details by the master equation
and Monte Carlo approaches.  Sequences $A$ and $B$ have been established
to be the good and bad folders respectively. Among the $15\;037$ different
conformations that a 12-monomer sequence can take, 495 are the local
energy minima for sequence $A$ and 496 for sequence $B$. The minima are
either $V$- or $U$-shaped.  The $U$-shaped minima are those in which a move
that does not change the energy is allowed, provided there are no
moves that lower the energy.  Both kinds of minima arise as a result
of the steepest descent mapping from states generated along a Monte
Carlo trajectory and both kinds are included in the disconnectivity
graphs.

Constructing a disconnectivity graph requires determination of the
energy barriers between each pair of the local energy minima. We do
this through an exact enumeration. We divide the energy scale into
discrete partitions of resolution $\Delta E$ (we consider $\Delta
E$=0.5) and ask between what minima there is a pathway which does not
exceed the threshold energy set at the top of the partition.  These
minima can then be grouped into clusters which are disconnected from
each other.  Local minima belonging to one cluster are connected by
pathways in which the corresponding barriers do not exceed a threshold
value of energy whereas the local minima that belong to different
clusters are separated by energy barriers which are higher than the
threshold level. At a sufficiently high value of the energy threshold
all minima belong to one cluster.  Enumeration of the pathways
involves storing a table of size $15\;037 \times 14$ because each
conformation may have up to 14 possible moves within the dynamics
considered in ref. \cite{Malte}.  (16-monomer heteropolymers can also be
studied in this exact way -- within any resolution $\Delta E$.)

Figure 1 shows the resulting disconnectivity graphs for sequence $A$.
For clarity, we show only this portion of the graph which involves the
local minima with energies which are smaller than $-5$ (there are 206
such minima). Throughout this paper, the symbol $E$ denotes energy
measured in terms of the coupling constants in the Hamiltonian and is
thus a dimensionless quantity. 
The native state, denoted as NAT in the Figure 1,
belongs to the most dominant valley.  One can see that the graph
contains a remarkable ``palm tree'' branch that provides a linkage to
the native state. This branch is a place within which a dynamically
defined folding funnel is expected to be confined to.  The large size
of this branch associated with a big energy gap between the native
state and other minima indicates large thermodynamic stability.  At low
temperatures, the glassy effects set in  and contributions due to
non-native valleys become significant.  The local minimum denoted by
TRAP in Figure 1 has been identified in ref. \cite{Malte} as giving
rise to the longest lasting relaxation processes in the limit of $T$
tending to 0.
 
The disconnectivity tree for sequence $B$ is shown in Figure 2.  Again,
only the minima with energies smaller than $-5$ are displayed (there are
203 such minima).  In this case, there are several local energy minima
which are bound to compete with the native state.  The corresponding
branches have comparable lengths and morphologies.  The dynamics is
thus expected not to be confined merely to the native basin.  Instead,
the system is bound to be frustrated in terms of what branch to choose
to evolve in.  At low $T$'s the valley containing the TRAP
conformation is responsible for the longest relaxation and poor
folding properties.

Other examples of disconnectivity trees for protein related systems have
been recently constructed with the use of Go-like models \cite{Li,Miller}
(in which the aminoacid-aminoacid interactions are restricted to
the native contacts)
and they confirm the general pattern of differences in morphology between
good and bad foldability as illustrated by Figures 1 and 2.

It should be noted that there are many ways to map out the multidimensional
energy landscape of proteins.  In particular, extensive energy landscape
explorations for the HP lattice heteropolymers have been done  with the use
of the pathway maps \cite{MillerDill,Chan1,Chan2}.  The pathway maps show
the actual microscopic paths through conformations. The paths are
enumerated either exactly or statistically, and thus provide a detailed but
implicit representation of the energy landscape. The resulting ``flow
diagrams'' indicate patterns of allowed kinetic connections between actual
conformations, together with the energy barriers involved. They can also be
additionally characterized by Monte Carlo determined probabilities to find
a given path at a temperature under study.  In this way, preferable
pathways and important transition states can be identified. This approach
is similar in spirit to the one undertaken by Leopold {\it et al.} 
\cite{Leopold}
in which the folding funnel is identified through determination of weights
associated with paths that lead to the native state.

The coarse grained representation of energy landscapes in protein-like
systems through the disconnectivity trees is quite distinct from that
obtained through the pathway maps.  The disconnectivity graphs indicate
only the one best path for each pair of the local energy minima by showing
the terminal points and the value of the energy barrier necessary to travel
this path.  This reduced information is precisely what allows one to
provide an explicit and essentially automatic visualization of the energy
landscapes.

\begin{figure}
\epsfxsize=3.2in
\centerline{\epsffile{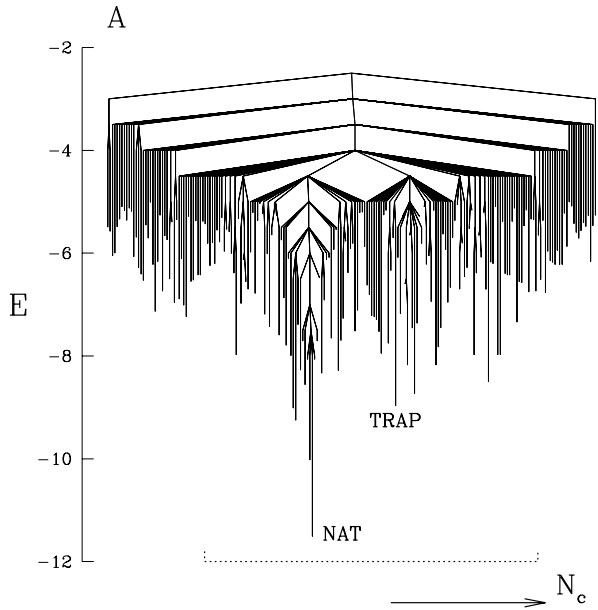}}
\caption{The disconnectivity graph for the 12-monomer
sequence $A$.  The dotted area is shown again in Figure 10 together with
the dynamical connectivities.  $N_c$ is a symbolic notation for a
label of an energy minimum, based on computer generated listing.}
\end{figure}

\begin{figure}
\epsfxsize=3.2in
\centerline{\epsffile{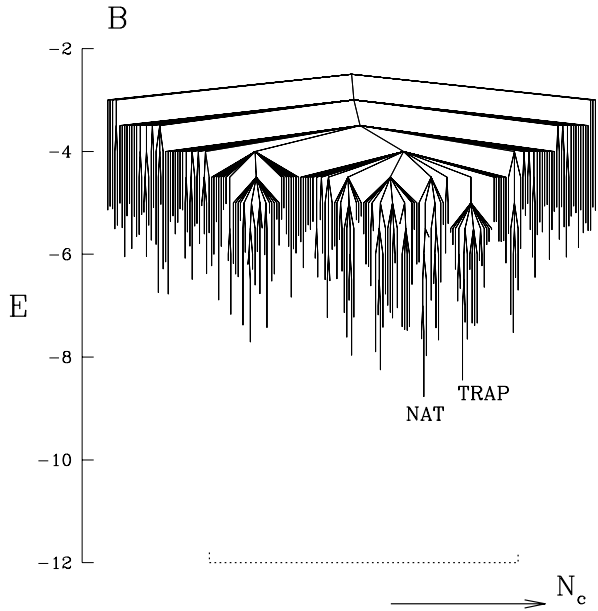}}
\caption{Similar to Figure 1 but for sequence $B$. The
dotted area is shown expanded in Figure 11. }
\end{figure}

The $T$-dependent frequencies of passages between conformations in the
pathway maps give an account of the dynamics in the system. This
information on the dynamics, however, does not easily fit the description
provided by the disconnectivity graphs. The steepest descent mapping to the
local energy minima that we propose here is, on the other hand, a perfect
match.

\section{ENERGY LANDSCAPES IN 2D SPIN SYSTEMS}

We now consider the spin systems. The Hamiltonian is given by $H\;=\;
\sum_{<ij>}J_{ij}S_i S_j$ where $S_i$ is $\pm1$, and the exchange
couplings, $J_{ij}$, connect nearest neighbors on the square lattice.
The periodic boundary conditions are adopted.  When studying spin
systems, a frequent question to ask about the dynamics is what are the
relaxation times -- characteristic times needed to establish
equilibrium.  Here, however, we are interested in quantities which are
analogous to those asked in studies of protein folding. Specifically,
what is the first passage time $t_0$? The first passage time is defined
as the time needed to come across the
ground state during a Monte Carlo evolution that starts from a random
spin configuration. A mean value of $t_0$ in a set of trajectories
(here, we consider 1000 trajectories for each $T$) will be denoted by
$\left<t_0\right>$ and the median value by $t_g$. $t_g$ is an analogue of the
folding time, $t_f$ of ref. \cite{Malte}.  At low temperatures, the
physics of relaxation and the physics of folding essentially agree
\cite{Malte}. At high temperatures, however, the relaxation is fast
but finding a ground state is slow due to a large entropy. Both for
heteropolymers and spin systems the $T$-dependence of the
characteristic first passage time is expected to be $U$-shaped.  The
fastest search for the ground state takes place at a temperature
$T_{min}$ at which the $T$-dependence has its minimum.

The $U$-shape dependence of $t_f$ originates in the idea of a low $T$
glassy phase in heteropolymers advocated by Bryngelson and 
Wolynes \cite{frustr} within the context of the random energy model.
It was subsequently confirmed in numerical simulations of 
lattice models \cite{MillerDill,Socci,Chan2}. This shape is, actually
expected for most disordered systems, including those involving spins.
However, experimentalists measuring spin systems typically would
not ask about the first passage time (at high $T$).

This overall behavior is illustrated in Figure 3 for two $5 \times 5$
spin systems. The Gaussian couplings of zero mean and unit dispersion are
selected for the spin glassy (SG) 2$D$ system. The disordered
ferromagnetic system (DFM) is endowed with the exchange couplings
which are the absolute values of the couplings considered for SG.
Figure 3 shows that $t_g$ does depend on $T$ in the $U$-shaped fashion.
$T_{min}$ for SG and DFM are comparable in values but the ``folding''
times for DFM are more than 4 times shorter than for SG. The times are
defined in terms of the number of Monte Carlo steps per spin.

\begin{figure}
\epsfxsize=3.2in
\centerline{\epsffile{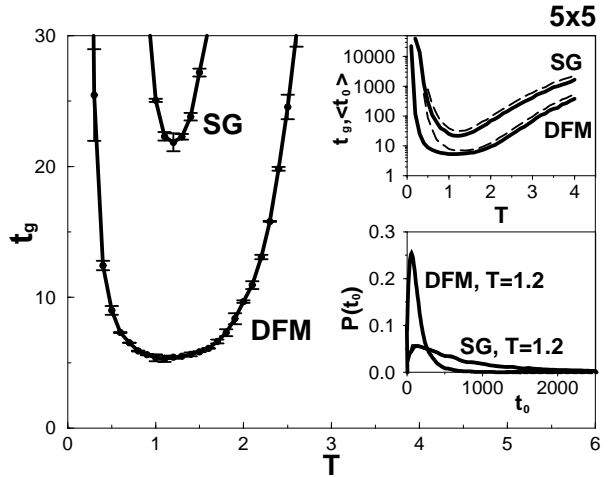}}
\caption{The main figure shows the $T$-dependence of
$t_g$ -- the median time to find the ground state -- for $5 \times 5$
DFM and SG systems.  The top inset compares $t_g$ to $t_0$ on the
logarithmic time scale.  The divergence of the two times at low $T$'s
indicates a substantial spreading out of the distribution of $t_0$.
This distribution, $P(t_0)$, is shown in the lower inset for
temperatures corresponding to $T_{min}$. }
\end{figure}

\begin{figure}
\epsfxsize=3.2in
\centerline{\epsffile{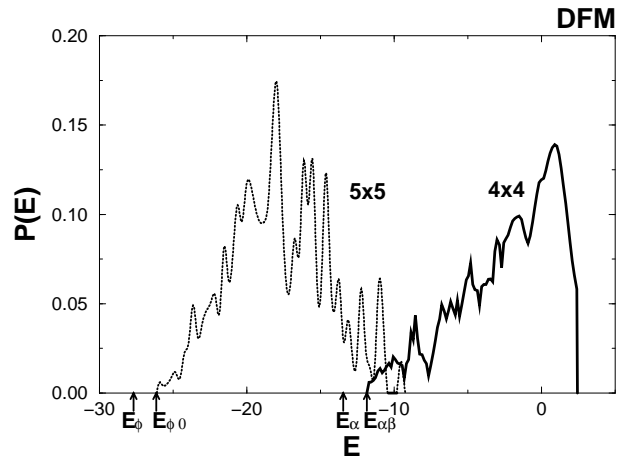}}
\caption{Distribution of energy barriers (highest elevation points) across
trajectories. The solid line is for the $4 \times 4$ DFM system.
It shows barriers for all trajectories which connect local minimum
$\alpha$ to another local minimum $\beta$. The lowest of them,
$E_{\alpha \beta}$, is used as threshold in the disconnectivity graph.
The dotted line is for a $5 \times 5$ DFM and for trajectories which
go from a local energy minimum $\phi$ to the ground state.
The energy barrier $E_{\phi 0}$ is obtained through the approximate
enumeration as described in the text. The other values are obtained
by generating 50000 random connecting trajectories. }
\end{figure}

Figure 3 establishes some of the analogies between the heteropolymers
and the spin systems. We now consider the disconnectivity graphs for
selected $L \times L$ spin systems with $L$=4 and 5.  For both system
sizes, the list of the local energy minima is obtained through an
exact enumeration. Determination of an energy barrier between two
minima requires adopting some approximations. Suppose that the two
minima differ by $n$ spins. There are then $n!$ possible trajectories
which connect the two minima, assuming that a) no spin is flipped more
than once, b) no other spins (or ``external'' spins) are involved in a
pathway.  These trajectories can be enumerated for $L$=4 but not for
$L$=5.  In the latter case we adopt the following additional
approximation.  We first identify the $n(n-1)(n-2)(n-3)$ list of
the first four possible steps in any trajectory together with the
highest energy elevation reached during these four steps. We choose
$m$=1500 trajectories which accomplish the smallest elevation. We then
consider the next two-step continuations of the selected trajectories
and among the $m(n-4)(n-5)$ continuations again select $m$ which
result in the lowest elevation, and so on until all $n$ spins are
inverted. The lowest elevation among the final set of the $m$
trajectories is an estimate of the energy threshold used in the
disconnectivity diagram.  This approximate method, when applied to the
$L$=4 systems, generates results which agree with the exact
enumeration.  Figure 4 shows that our method clearly beats
determination of barriers based on totally random trajectories (but
still restricted to overturning of the $n$ differing spins).

Flipping of the ``external'' spins was found to give rise to an
occasional reduction in the barrier height. We could not, however,
come up with a systematic inclusion of such phenomena in the
calculations and the resulting disconnectivity graphs have barriers
which are meant to be estimates from above. The topology of the graph
is expected to depend little on details of such approximations.

\begin{figure}
\epsfxsize=3.2in
\centerline{\epsffile{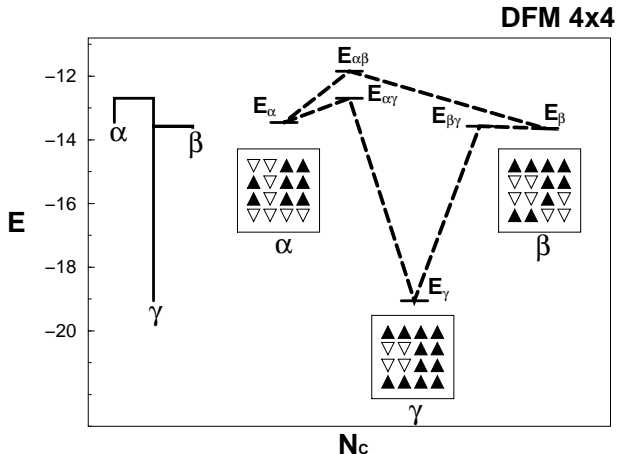}}
\caption{Examples of pathways between three local energy
minima, $\alpha, \beta$, and $\gamma$, in a $4 \times 4$ DFM. The
corresponding spin configurations are shown by arrows. The resulting
disconnectivity graph is shown on the left. }
\end{figure}

In some cases, the barrier for a direct travel from one minimum to
another was found to be higher than when making a similar passage via
an intermediate local energy minimum.  An example of this situation is
shown in Figure 5.  However, this lack of transitivity, resulting from
the approximate nature of the calculations, does not affect the
disconnectivity graph because the states $\gamma$ and $\beta$ of
Figure 5 are mutually accessible at energy $E_{\beta\gamma}$. Then, at
a higher energy $E_{\alpha\gamma}$, state $\alpha$ is thus also
accessible. If, at this energy level, the system can transfer between
the states $\alpha$ and $\gamma$ then it can also transfer to state
$\beta$.

\begin{figure}
\epsfxsize=3.2in
\centerline{\epsffile{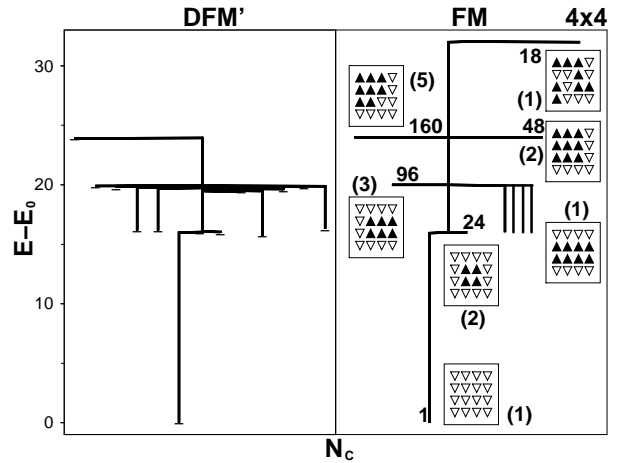}}
\caption{The disconnectivity graphs for the FM and DFM'
$4\times 4$ systems, as defined in the text. The arrows in the boxes
show examples of the corresponding spin configurations. The numbers
indicate the degree of degeneracy and the numbers in brackets indicate
numbers of distinct geometries for the inverted domains to take at the
energy considered.  For a uniform antiferromagnetic system the
disconnectivity graph looks qualitatively similar to the one
characterizing FM but the ground state is the only $V$-shaped local
energy minimum of the system.}
\end{figure}

We now present specific examples of disconnectivity graphs for several
distinct spin systems.

Figure 6 shows the case of a $4\times 4$ uniform ferromagnet (FM).
The energy landscape of the FM is not analogous to that of a protein
because uniform exchange couplings generate states with high degrees
of degeneracies. These degeneracies can be split either by a
randomization.  Figure 6 also shows a graph for a $L=4$ DFM' system in
which the $J_{ij}$'s are random numbers from the [0.9,1.1] interval --
this is the case of a small perturbation away from the uniform FM.
The graph for DFM' has an overall appearance like the one for FM
except for the lack of a high energy linkage to a set of state which
cease to be minima. Another difference is the disappearance of all
remaining $U$-shaped minima and formation of new true minima at somewhat
spread out energies.  In the uniform $L=4$ ferromagnet, there are five
$V$-shaped energy minima: one is the ground state and the other four
higher energy states are degenerate.  In addition, there are 346
states which are the $U$-shaped energy minima.  An example of what
happens in a $U$-shaped minimum is shown in Figure 7.  Here, the system
can move between the 3- and 4-spin domains without a change in the
energy. The 4-spin domain forms a $U$-shaped minimum but the 3-spin
state is not a minimum because there is a move to a lower energy
state. Only the 4-spin domain states would be shown in the
disconnectivity graph. 

\begin{figure}
\epsfxsize=3.2in
\centerline{\epsffile{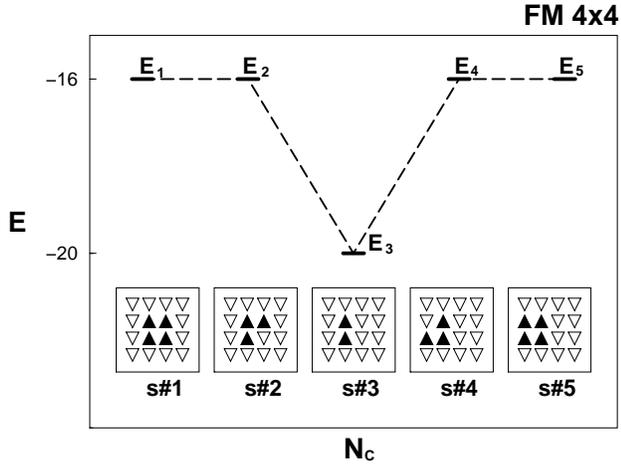}}
\caption{Examples of spin configurations in the $4\times 4$ FM system.}
\end{figure}

\begin{figure}
\epsfxsize=3.2in
\centerline{\epsffile{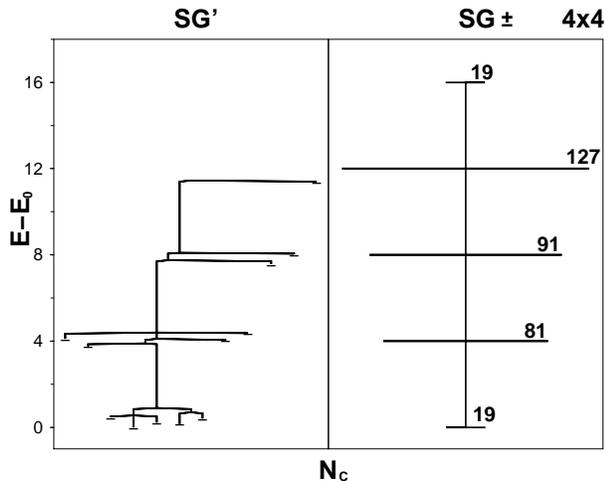}}
\caption{The disconnectivity graphs for the $4 \times4$
spin glassy systems: with the Gaussian (SG') and with the $\pm 1$
couplings (SG $\pm$).  The numbers correspond to the number of
$U$-shaped minima at an energy  shown in the graph.}
\end{figure}

Figure 8 shows the disconnectivity graphs for two $L=4$ spin glassy
systems. The right hand panel shows the case of $J_{ij}= \pm 1$.  The
left hand panel shows a spin glass (SG') with the exchange couplings
which are randomly positive or negative and with their magnitudes
coming from the interval [0.9,1.1] -- this is the random sign
counterpart of the DFM' system.  In both spin glassy systems of Figure
8 the allocation of signs to the couplings is identical.  In the $\pm
1$ case, all minima, including the degenerate ground state, are
$U$-shaped. The SG' system, on the other hand, has a graph with an
overall structure akin to that corresponding to the $\pm 1$ system
with one important difference: the ground state  is not degenerate and
thus the ground state basin splits into several competing valleys.

The differences between the good and bad spin ``folders'' amplify as the
system size is increased. As an illustration, Figure 9 shows the
disconnectivity graphs for the $5 \times 5$ DFM and SG -- with the
Gaussian couplings.  The DFM systems has a very stable and well
developed valley corresponding to the ground state whereas the SG
system has many competing valleys. Thus indeed, DFM is a spin analogue
of a protein whereas SG is an analogue of a random sequence of
aminoacids.

\begin{figure}
\epsfxsize=3.2in
\centerline{\epsffile{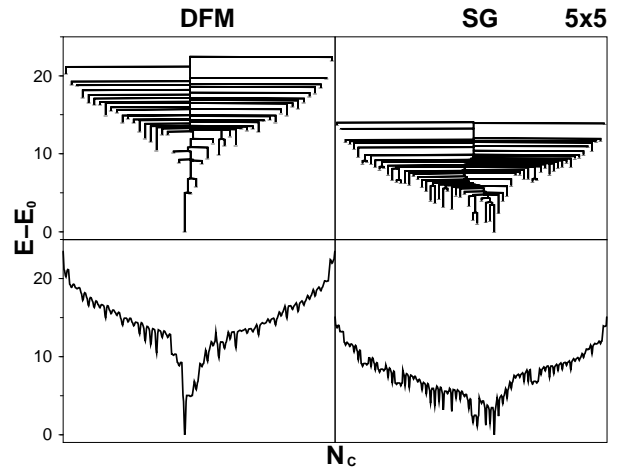}}
\caption{The disconnectivity graphs for the $5 \times 5$
DFM and SG systems (the top panels) and the corresponding
representation of the energy landscapes (the bottom panels).}
\end{figure}

The disconnectivity graphs can be represented in a form that gives a
better illusion of an actual landscape, as shown in the bottom panels
of Figure 9. The lines shown there connect the local energy minima to
their energy barriers and then to the next minimum, and so on, forming
an envelope of the original graph. This form is less cluttered and
will be used in Sec. V. This envelope representation shows merely
the smallest scale variations in energy and omits passages with
large barriers.

\section{DYNAMICAL CONNECTIVITY GRAPHS FOR LATTICE HETEROPOLYMERS}

We now construct the supergraphs for the lattice heteropolymers
discussed in Sec. II. The strengths of the dynamical linkages have been
already determined in ref. \cite{coarse} at several temperatures.
Here, however, we plot the linkages on the graphs that represent the
energy landscapes, i.e. we rearrange the labels associated with the
local energy minima. We discuss only the case of $T=T_{min}$ which is
equal to 1.0 for both sequences $A$ and $B$.

\begin{figure}
\epsfxsize=3.2in
\centerline{\epsffile{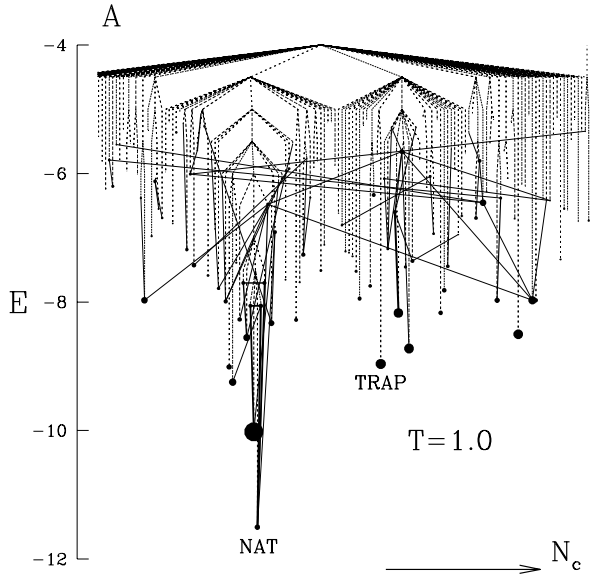}}
\caption{Dynamical connectivity graph for sequence $A$ at
$T$=1.0 plotted against the background of the disconnectivity graph.
The dynamical linkages are restricted to the dotted region of Figure 1
and only this portion of the disconnectivity graph is shown.}
\end{figure}

\begin{figure}
\epsfxsize=3.2in
\centerline{\epsffile{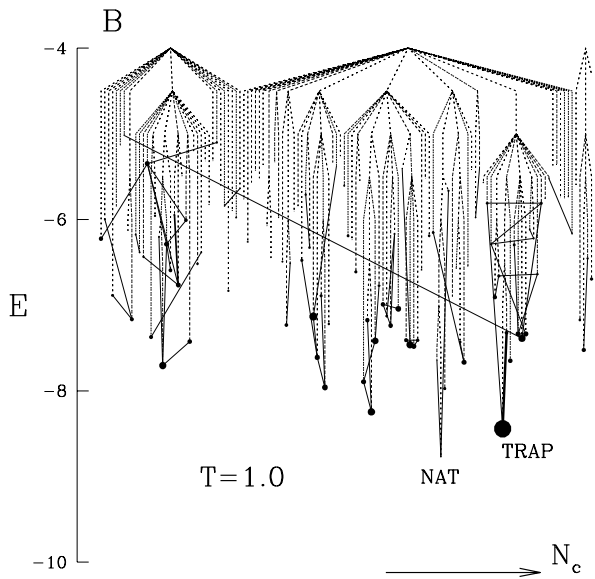}}
\caption{Similar to Figure 10 but for sequence $B$.}
\end{figure}

Figures 10 and 11 shows the supergraphs for sequences $A$ and $B$
respectively.  The sizes of the circles are proportional to an
occupancy of the minimum during the folding time. Similarly, the
thicknesses of the lines connecting the circles are proportional to
the connectivity (the linking frequency) between them.  For clarity,
we do not show connectivities which account for less than $1\%$ of all
combined dynamical connectivities. The disconnectivity graphs
themselves are drawn in dotted lines.  All relevant dynamics is
confined to these portion of the of the disconnectivity graphs which
were marked, in Figures 1 and 2, by the dotted lines and are now
magnified in Figures 10 and 11.

An inspection of the supergraphs clearly shows differences between the
two sequences. Sequence $A$ has many inter-valley linkages but the
linkages to the native basin, and the occupancies of conformations
within that basin, are substantial.  These are manifestations of a
fast folding dynamics. For sequence $B$, on the other hand, the linkages
tend to wither uncooperatively in multiple valleys. In addition, the
combined occupancies away from the native valley outweigh the
dynamical effects within the valley.  On lowering the temperature,
linkages in various valleys become disconnected and tend to  avoid the
native valley more and more, as discussed in ref. \cite{coarse}.

\section{DYNAMICAL CONNECTIVITY GRAPHS FOR SPIN SYSTEMS}

We now generate dynamical linkages for two spin systems, $L=5$ DFM and
SG of Sec. III, and place them on the plots of the energy landscape.
The ``envelope'' form of the representation of the landscape is chosen
here, mostly for esthetic reasons.  The connectivities are determined
based on 200 Monte Carlo trajectories of a fixed length of 5000 steps
per spin. The duration of these trajectories exceeds the ``folding
time'' many times, at the temperatures studied, and thus the
connectivities displayed  refer to the essentially equilibrium
situations (the equilibrium dynamics for heteropolymers $A$ and $B$ is
illustrated in ref. \cite{coarse}).  The connectivity rates were
updated any time (in terms of single spin events and not in terms of
steps per spin) there is a transition from a local energy minimum to a
local energy minimum, after the steepest descent mapping.  Again, the
$1\%$ display cutoff has been implemented when making the figure.

The main parts of Figures 12 and 13 show the supergraphs obtained at a
temperature which corresponds to the $T$-location of the peak in
specific heat. These temperatures, 1.8 for DFM and 1.4 for SG, are
also close to $T_{min}$. The insets show the dynamically relevant
parts of the energy landscape at lower temperatures.  For the DFM, the
dynamics becomes increasingly  confined to the ground state basin when
the temperature is reduced.  On the other hand, for the SG, the
dynamics in the ground state basin becomes less and less relevant,
with a higher local energy minimum absorbing the majority of moves.
This is indeed what happens with bad folding heteropolymers.

If we restrict counting of the transition rates only to the ``folding
stage,'' i.e. till the ground state is encountered, the qualitative
look of the supergraph for $T$ close to $T_{min}$ is as in the
equilibrium case. The states involved are mostly the same but there
is, by definition, only one link to the ground state per trajectory.  

The dynamical connectivity graphs in 3D $10\times 10\times 10$ DFM
systems are qualitatively similar to the 2D graphs but the
underlying disconnectivity graphs are harder to display due to a
substantially larger number of the energy minima.

\begin{figure}
\epsfxsize=3.2in
\centerline{\epsffile{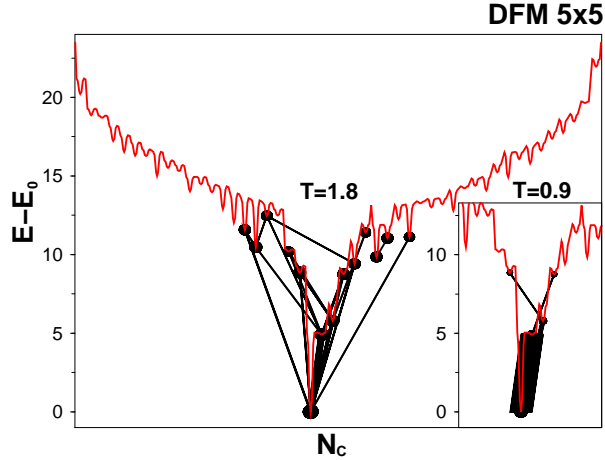}}
\caption{The dynamical-equilibrium supergraph for the
$5\times 5$ DFM system at $T$=1.8. The inset shows the portion
that is relevant at $T$=0.9.}
\end{figure}

\begin{figure}
\epsfxsize=3.2in
\centerline{\epsffile{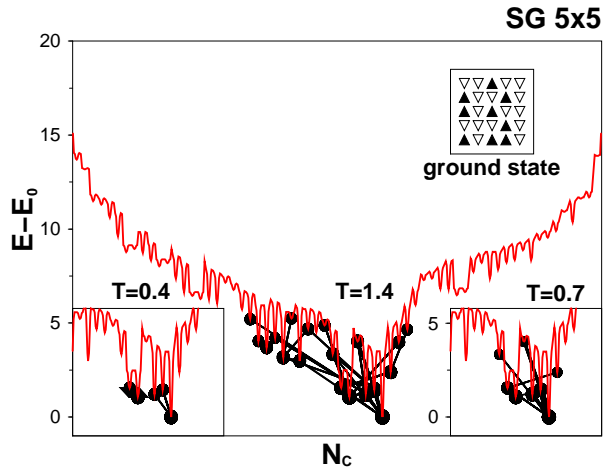}}
\caption{The supergraph for the $5\times 5$ SG system at
$T$=1.4. The insets show the portions which are relevant at two lower
temperatures. The ground state configuration is shown at the top.}
\end{figure}


In this paper we have pointed out the existence of many analogies
between protein folding and dynamics of spin systems.  These analogies
have restrictions. For instance the simple Ising spin systems in 3$D$
have continuous phase transitions, in the thermodynamic limit, and not
the first-order-like that are expected to characterize large proteins
\cite{Gutin}.  This difference, however, is not crucial in the case of
small systems.  More accurate spin analogs of proteins, with the first
order transition, can be constructed but the object of this paper was
to discuss the basic types of spin systems.

On the other hand, it should be pointed out that these analogies are
also more extensive.  Consider, for instance, the Thirumalai
\cite{Thirumalai} criterion for good foldability of proteins. The
criterion considers two quantities: the specific heat and the
structural susceptibility of a heteropolymer.  The latter is a measure
of fluctuations in the structural deviations away from the native
state. Both quantities have peaks at certain temperatures. The
criterion specifies that if the two temperatures coincide a
heteropolymer is a good folder. This is quite similar to what happens
in uniform and disordered 3D ferromagnets: the peaks (singularities)
in magnetic susceptibility and specific heat are located  at the same
critical temperature. On the other hand, in spin glasses, the broad
maximum in the specific heat is located at a temperature which is
substantially above the freezing temperature associated with the cusp
in the susceptibility. Also in this sense then, spin glasses behave
like bad folders.

The coarse-graining supergraphs that analyse dynamics in the context
of the system's energy landscape may become a valuable tool to
understand  complex behavior of many body systems.

\section*{ACKNOWLEDGMENTS}

This work was supported by KBN (Grants No. 2P03B-025-13 and
2P03B-125-16).  Fruitful discussions with Jayanth R. Banavar are
appreciated.

\end{document}